# Quantum Limited DPSK Receivers with Optical Mach-Zehnder Interferometer Demodulation

**Xiupu Zhang**,
*Department of Electrical and Computer Engineering,
Concordia University, Montreal, Quebec, CANADA,
E-mail: xzhang@ece.concordia.ca)*

**Guodong Zhang**,
*AT&T, 200 Laurel Avenue, Middletown, NJ 07748, USA*

*Abstract:* We present an analysis of quantum-limited DPSK receivers with optical Mach-Zehnder interferometer (MZI) demodulation. It is shown for the first time that the quantum limits for DPSK/MZI receivers with single-port and balanced detections exactly differ from 3-dB in receiver sensitivity, obtained by both Poisson and Gaussian noise statistics. The quantum limit for DPSK/MZI receivers with balanced detection is given by $BER = \frac{1}{2}\exp\left(-2\bar{N}_p\right)$ for the first time, instead of $BER = \frac{1}{2}\exp\left(-\bar{N}_p\right)$ which only applies for DPSK/MZI receivers with single-port detection, $\bar{N}_p$ - the photon number in bit "1" or "0", i.e. average photon number.

## 1. Introduction

The differential phase shifted keying (DPSK) modulation has been attracted great attention for its application for dense wavelength division multiplexing (DWDM) transmission since DPSK with optical Mach-Zehnder interferometer (MZI) demodulation and balanced detection provides several advantages over the conventional intensity modulation/direction detection (IM/DD) [1]. The quantum limited receiver sensitivity of DPSK receivers, determined by $BER = \frac{1}{2}\exp\left(-\bar{N}_p\right)$ - the photon number in bit "1" or "0", i.e. *average photon number* (bits "1" and "0" carry the same signal energy in DPSK signal) and BER- bit error ratio, has been widely used for DPSK/MZI receivers with both single-port and balanced detections [1-2]. The above quantum limit was obtained for DPSK with electrical demodulation (referred to the conventional DPSK receivers), which consists of an electrical time delay line and a mixer, based on the noise statistics of Rice (bit "1") and Rayleigh (bit "0") distributions [3-4]. However, the MZI (in DSPK with optical MZI demodulation) converts DPSK optical signal into intensity modulated optical signal before input to the optical photodiodes, which is shown in Fig. 1. Consequently the electrical processing of DPSK signal/noise in optical receivers is the same as in IM/DD receivers, rather than the conventional DPSK receivers. The noise statistic of quantum noise (i.e. shot noise) in DPSK receivers with optical MZI demodulation is not the Rice and Rayleigh probability distributions; instead the **Gaussian/Poisson** noise distribution should be used as in IM/DD receivers [5]. Moreover, the DPSK/MZI receivers with balanced detection could be different from DPSK/MZI receivers with single-port detection in quantum limited receiver sensitivity, because the signal energy used for error detection is different in the two detections. Consequently, it could be expected that the quantum limited BER in DPSK/MZI receivers with single-port and balanced detections may be different from that of the conventional DPSK receivers. In this paper, we present a quantum limited analysis for DPSK receiver with optical MZI demodulation and single-port and balanced detections.

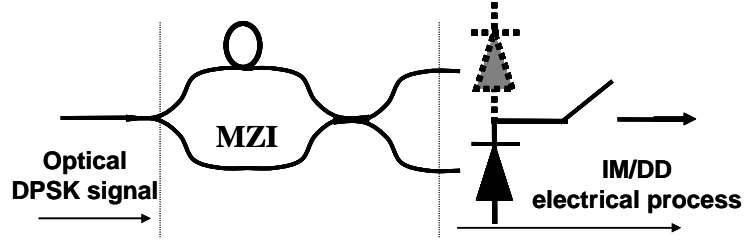

Fig. 1 Schematic drawing of a DPSK/MZI receiver. Single-port detection uses one photodiode and balanced detection both photodiodes. MZI is used for conversion from phase modulation to intensity modulation.

## 2. Definitions of quantum and quasi-quantum noise

When quantum noise is only considered, a small number of photons and electron-hole pairs present (i.e., the number of photons and electrons are countable). The noise statistics (only quantum noise is taken into account) for DPSK/MZI receivers should follow the Poisson distribution (a discrete probability distribution) as in IM/DD receivers [5]. As the number of photons and electrons becomes large enough, the noise statistics become the Gaussian distribution (a continuous probability distribution). In this paper, quasi-quantum limited (QQL) analysis is referred if the quantum noise is considered to be the Gaussian noise, to distinguish it from the quantum limited (QL) analysis in which the quantum noise is considered to be Poisson noise. For the conventional DPSK receivers, the BER expression of $BER = \frac{1}{2}\exp(-\bar{N}_p)$ [3-5] is corresponding to our defined quasi-quantum limited analysis because the continuous Rice and Rayleigh noise statistics are used.

## 3. Quantum limited analysis

We first analyze the quantum limited (Poisson noise statistics) DPSK receivers with optical MZI demodulation. We first consider DPSK/MZI receivers with single-port detection. If $m_1 > 0$ electron-hole pairs with the Poisson probability of $P(m_1) = \exp[-\bar{N}_p]\bar{N}_p^{m_1}/m_1!$ are generated by photon number $\bar{N}_p$ ($\bar{N}_p$ - the photon number in bit "1", and corresponding to the average optical power of the DPSK signal), no errors from bit "1" occur. Since bit "0" has zero photons and noise free, bit "0" is not detectable and BER is totally determined by bit "1" similar to IM/DD receivers [5] (Note $m_0 \equiv 0$ the number of electron-hole pairs in bit "0"). Therefore the quantum limited BER is given by setting $m_1 = 0$ in the above Poisson distribution, i.e.

$$BER_{S-QL} = \frac{1}{2}\left[\exp(-\bar{N}_p) + 0\right] \qquad (1).$$

The receiver sensitivity given by (1) is 3-dB worse than that in IM/DD receivers [5] ($BER_{IM/DD} = \frac{1}{2}\exp(-2\bar{N}_p)$, the peak power of bit "1" in IM/DD is assumed twice the average power of DPSK signal and thus total signal energy carried by IM and DPSK signals is the same). This can be explained that only the half signal energy is used for error detection in DPSK/MZI receivers with single-port detection rather than the full signal energy in IM/DD receivers. The result indicated by (1) is already given in [5, Table 10.2] for the conventional DPSK receivers.

For DPSK/MZI receivers with balanced detection, the bits "1" and "0" contain the same number of photons. When bit "1" transmitted, no errors occur if $m_1 > 0$ electron-hole pairs with probability of $P(m_1) = \exp[-\bar{N}_p]\bar{N}_p^{m_1}/m_1!$ are generated at the constructive port. Similarly, no errors occur from bit "0" if $m_0 > 0$ electron-hole pairs with probability of $P(m_0) = \exp[-\bar{N}_p]\bar{N}_p^{m_0}/m_0!$ are created at the destructive port. Thus, no errors occur if the condition $m_0 + m_1 > 0$ is met by combining the two conditions. For example, we consider a special case: $m_1 > 0$ and $m_0 = 0$. This case is exactly the same as DPSK/MZI receivers with single-port detection, in which bit "1" has $m_1 > 0$ and bit "0" $m_0 = 0$ and thus no errors occur. In other words, it was shown above that no errors occur if $m_1 > 0$ and $m_0 = 0$, and vice versa. Thus, an error shall occur only if $m_0 + m_1 = 0$ with the probability of $P(m = m_0 + m_1) = \exp[-2\bar{N}_p](2\bar{N}_p)^m/m!$. Thus, for DPSK/MZI receivers with balanced detection, BER is given by setting $m = 0$,

$$BER_{B-QL} = \frac{1}{2}\exp[-2\bar{N}_p] \qquad (2).$$

The factor 1/2 is due to two bits. By comparing (1) and (2), we can find that the 3-dB quantum limited receiver sensitivity is improved by DPSKMZI receivers with balanced detection over single-port detection. On the other hand, the same quantum limit by DPSK/MZI receivers with balanced detection as IM/DD receivers ($BER_{IM/DD} = \frac{1}{2}\exp(-2\bar{N}_p)$ since bit "1" $N_p = 2\bar{N}_p$ in IM/DD receivers) is obtained. This is because the two receivers use the same signal energy for error detection. The expression (2) for DPSK/MZI receivers with balanced detection is given for the first time. It is shown that the quantum limits are different for DPSK/MZI receivers with single-port and balanced detections. Therefore, it is not appropriate to use the expression (1) for DPSK/MZI receivers with balanced detection [1-2]. If non-ideal photodiodes are considered, $\eta\bar{N}_p$ should replace $\bar{N}_p$ in (1) and (2), $\eta$ - the quantum efficiency of the photodiodes.

Furthermore, it is observed that the expression (1) for DPSK/MZI receivers with single-port detection is the same as the conventional DPSK receivers. However the BER given by (1) is obtained based on the discrete Poisson distribution, rather than the continuous Rice and Rayleigh distributions. Particularly, it is worth to emphasize that the quantum-limited BER expression of $BER = \frac{1}{2}\exp(-\eta\bar{N}_p)$, which has been widely used for DPSK receivers [1-2], only applies for DPSK/MZI receivers with single-port detection and the conventional DPSK receivers. Additionally, the 3-dB difference of receiver sensitivity by (1) and (2) agrees well with the signal constellation which is shown in Fig. 2 [1]. In Fig. 2(a), the signal constellations for DPSK/MZI receivers with single-port detection and IM/DD receivers are given. The distance between bits "1" and "0" in electric field is assumed $x$ for DPSK/MZI receivers with single-port detection. Thus, the distance becomes $\sqrt{2}x$ for IM/DD receivers. Therefore, IM/DD receivers outperform DPSK/MZI receivers with the single-port detection by 3 dB. Fig. 2(b) depicts the signal constellations for IM/DD receivers and DPSK/MZI receivers with balanced detection. It is shown that the distances between bits "1" and "0" are the same for the two receivers. Therefore, IM/DD receivers have the same quantum limited performance as DPSK/MZI receivers with balanced detection.

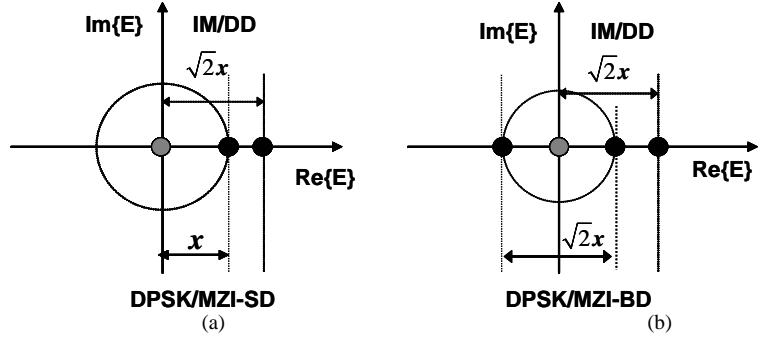

Fig.2 (a) signal constellation for DPSK/MZI receivers with the single-port detection (DPSK/MZI-SD) and IM/DD receivers, (b) signal constellation for DPSK/MZI receivers with the balanced detection (DPSK/MZI-BD) and IM/DD receivers.

**4. Quasi-quantum limited analysis**

We now start the analysis for the quasi-quantum limited (Gaussian noise statistics) DPSK receivers with optical MZI demodulation. First let's consider the DPSK/MZI receivers with single-port detection. The decision current $I_1(t)$ for bit "1" is corresponding to the average optical power rather than the peak power in IM/DD receivers. The decision currents for DPSK/MZI receivers with constructive-port detection are $I_1(t) = R\overline{P}_s + n_s(t)$ for bit "1" and $I_0(t) = 0$ for bit "0", where R is the responsivity of the photodiodes, $\overline{P}_s$ denotes the average optical power, and $n_s(t)$ is the quantum noise with the variance of $\sigma^2$. The quasi-quantum limited BER for DPSK/MZI receivers with single-port detection is similar to IM/DD [5],

$$BER_{S-QQL} = \frac{1}{2}erfc\left[\frac{I_s}{\sigma\sqrt{2}}\right] = \frac{1}{2}erfc\left(\sqrt{\frac{\eta\overline{N}_p}{2}}\right) \quad (3),$$

where $erfc()$ is the complementary error function. In (3) $I_s = R\overline{P}_s$, and $\sigma^2 = 2eI_s B_e$, i.e. the shot noise for bits "1", $e$ - electron charge, $B_e$ - the electrical noise bandwidth. For $B_e$ equal to the half of the bit rate, we obtain $\frac{I_s^2}{\sigma^2} = \eta\overline{N}_p$, which is used in the last step of (3). It is seen that DPSK/MZI receivers with single–port detection is 3-dB worse than IM/DD receivers in receiver sensitivity ( $BER_{IM/DD} = \frac{1}{2}erfc\left[\sqrt{\frac{\eta 2\overline{N}_p}{2}}\right]$ [5]), again the same conclusion as the quantum-limited analysis.

For DPSK/MZI receivers with balanced detection, the decision currents are $I_1(t) = R\overline{P}_s + n_s(t)$ for bit "1" and $I_0(t) = -R\overline{P}_s + n_s(t)$ for bit "0". The quantum noise $n_s(t)$ of bits "1" and "0" is the same in the variance with $\sigma^2 = 2eI_s B_e$. By combining the two decision conditions as the case of quantum limit, we have the error occurring condition of $I_1(t) - I_0(t) < 0$ [6]. BER can be obtained by,

$$BER_{B-QQL} = \Pr ob(I_1 < I_0) = \int_{-\infty}^{\infty} \frac{1}{\sqrt{2\pi}\sigma} \exp\left[-\frac{(x-I_s)^2}{2\sigma^2}\right] dx \int_{x}^{\infty} \frac{1}{\sqrt{2\pi}\sigma} \exp\left[-\frac{(y+I_s)^2}{2\sigma^2}\right] dy$$

$$= \frac{1}{2} erfc\left(\frac{I_s}{\sigma}\right) = \frac{1}{2} erfc\left(\sqrt{\eta \overline{N}_p}\right) \qquad (4).$$

In (4) the same conditions as in (3) have been applied in the last step. By comparing (3) and (4), we have found that BER given by (3) and (4) differs from 3-dB in receiver sensitivity. In other words, the 3-dB receiver sensitivity is improved by DPSK/MZI receivers with balanced detection over single-port detection in the quasi-quantum limit. On the other hand, DPSK/MZI receivers with balanced detection has the same quantum limit as IM/DD receivers, since the total signal energy, used for error detection in DPSK/MZI receivers with balanced detection, is exactly the same as in IM/DD receivers. The BER expressions of (3) and (4) are different from the expression of $BER = \frac{1}{2}\exp(-\eta \overline{N}_p)$ obtained for the conventional DPSK receivers based on the continuous Rice and Rayleigh distributions. Again the 3-dB receiver sensitivity difference in (3) and (4) can be easily interpreted by the signal constellation in Fig. 2.

## 5. Conclusion

We have presented an analysis of DPSK/MZI receivers with single-port and balanced detections, considering the quantum noise only. We have found that 3-dB quantum limit is improved by DPSK/MZI receivers with balanced detection over single-port detection. This is simply because only the half signal energy is used for error detection in single-port detection. Moreover, DPSK/MZI receivers with balanced detection has the same quantum limit as IM/DD receivers rather than 3-dB lower, since the total signal energy for error detection in DPSK/MZI receivers with balanced detection and IM/DD receivers is the same. The quantum limited BER with $BER_s = \frac{1}{2}\exp(-\eta \overline{N}_p)$ for DPSK/MZI receivers with single-port detection and $BER_b = \frac{1}{2}\exp(-2\eta \overline{N}_p)$ for balanced detection are given for the first time, based on the Poisson statistic.

**Acknowledgements:** The authors thank Chongjin Xie, Bell Labs. Lucent Technologies, for reading the manuscript and suggestions.